\documentclass{article}

\PassOptionsToPackage{numbers, compress}{natbib}

\usepackage[preprint]{neurips_2022}

\usepackage[utf8]{inputenc} 
\usepackage[T1]{fontenc}    
\usepackage{hyperref}       
\usepackage{url}            
\usepackage{booktabs}       
\usepackage{amsfonts}       
\usepackage{nicefrac}       
\usepackage{microtype}      
\usepackage{xcolor}         
\usepackage{graphicx}

\usepackage{amsmath}
\DeclareMathOperator*{\argmaxB}{argmax} 

\title{A Learned Simulation Environment to Model Student Engagement and Retention in Automated Online Courses}

\author{
N. Imstepf $^1$, S. Senn $^1$, A. Fortin $^2$, B. Russell $^2$, and C. Horn $^1$ \\
$^1$ Zurich University of Applied Sciences, ICLS, Schloss 1, 8820 Wädenswil, Switzerland \\
$^2$ Kikodo Education Technology Sàrl, Av. Louis Ruchonnet, Lausanne 1003, Switzerland 
}

\begin{document}

\maketitle

\begin{abstract}
We developed a simulator to quantify the effect of exercise ordering on both student engagement and retention. 
Our approach combines the construction of neural network representations for users and exercises using a dynamic matrix factorization method.
We further created a machine learning models of success and dropout prediction.  
As a result, our system is able to predict student engagement and retention based on a given sequence of exercises selected.
This opens the door to the development of versatile reinforcement learning agents which can substitute the role of private tutoring in exam preparation.
\end{abstract}


\section{Introduction}

Survey data reveals that 50\% of US and UK households would like to hire a tutor, but cannot afford it \cite{montacute_2018}.
This issue is especially acute in the context of high-stakes testing such as GCSE and SAT/ACT which can have a determining effect on life outcomes of students, while results in such tests are also strongly influenced by the ability to hire a quality private tutor \cite{ireson_2004_private,moore_2018_investigating}.

Furthermore, a major concern for sustained economic growth today is the shortage of skilled workers \cite{bessen_2014_employers}.
Online learning combined with new approaches in education technology may help solve this challenge by improving access to the highest quality independent learning materials and processes.
Global AI usage in education is estimated to have an annual growth rate of 36.6 percent during 2022-2030 and is predicted to reach \$25.77 billion by 2030 \cite{psmarketresearch_2022}.

Classical classroom teaching often suffers from large percentages of disengaged students for which the presented material is either too difficult or too easy \cite{Khan}.
This problem is exacerbated by the diversity of the existing knowledge of the class.
Thus, the promise of personalized learning is to optimize the learning path for each student individually, and therefore maximize learning, retention, and ultimately exam performance for each student.

While past AI-based approaches have used models trained by supervised learning, we propose to use intelligent agents which are able to continuously learn throughout the complete student journey.
Training such agents requires the development of a simulator (aka digital twin) for student actions, which is the topic of this paper.
It forms the basis for a natural next step, the development of an autonomous digital tutor.

Our main contributions are: 
\begin{itemize}
    \item 
    An end-to-end approach to optimizing student interactions in online learning;
    \item 
    A novel dynamic matrix factorization method to learn neural representations of users and exercises from user data; 
    \item 
    A demonstration of how Model-Based reinforcement learning can be enabled by learning the components of a simulator.
\end{itemize}

The benefits of this approach translate into 
a more engaging learning experience for the students, higher performance, and increased retention rates. 

\section{Method}
Our aim is to build a simulator that allows us to quantify the effect of different exercise orders on student engagement and retention.
This could then be used to understand and improve the individual learning path of each student. In particular, such an approach provides the basis for the development of autonomous, intelligent agents to be deployed in automatic online courses. Using these agents, one can dynamically optimize student engagement and retention as a function of past interactions.
\begin{figure}
  \begin{center}
    \includegraphics[width=0.95\textwidth]{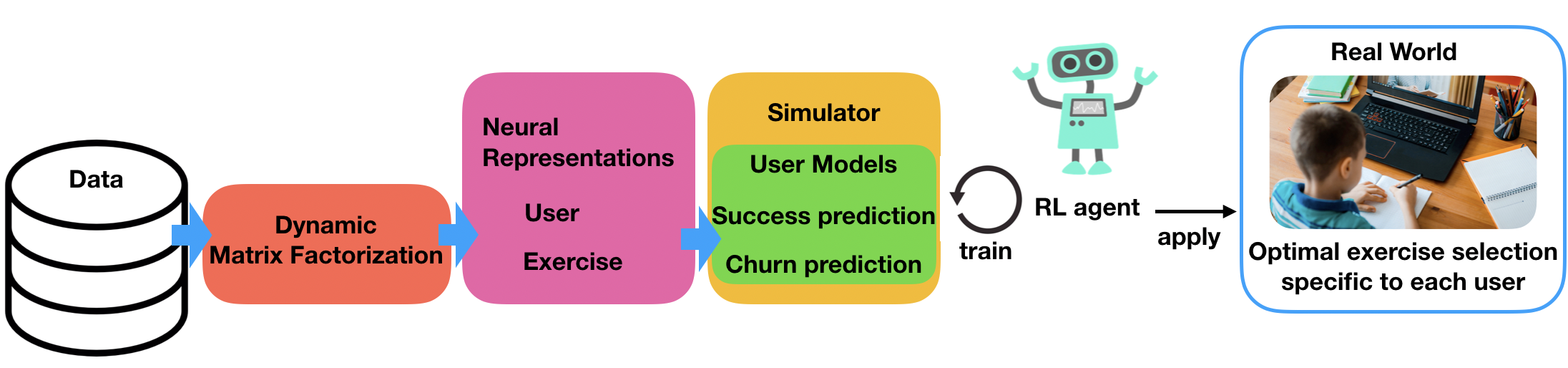}
  \end{center}
  \caption{Overview of our proposed modeling chain.}
  \label{fig:modelingChain}
\end{figure}
The corresponding modeling chain is illustrated in Fig. \ref{fig:modelingChain}. It comprises four main steps, (1) data gathering and pre-processing, (2) learning of neural representations for users and exercises, (3) modeling of success and dropout probability (4) construction of the student interaction simulator. All steps are described in detail below. 

\subsection{Collection and processing of student interaction data}

The user question submission data was based on the Python learn-to-code site www.kikodo.io.
For each user, a time series of "events" was used, where each event is the submission of a single exercise, and the outcome of that submission.
Although additional metadata were present in the original data, only the correctness or incorrectness of the submission was considered.

The exercises on this specific e-learning site are structured simply into "workbooks", which are themselves grouped by topic.
Each workbook contains a number of questions in the range of 10-100.
Prior to the preparation of this study all learners had questions within a workbook presented to them in a random order.

The users of the site have been anonymized in all reporting and analysis.

\subsection{Dynamic matrix factorization}   
\label{sec:MF}
Developing a neural network model to select the best exercise for a given user 
requires representations of users and exercises.
A common way to calculate these from user-exercise interaction data is collaborative filtering (CF) as used in recommender systems \cite{ricci_2011}.
While standard user-based and item-based CF just take the user-item\footnote{In our case the items are the exercises.} matrix to derive representations, matrix factorization actively optimizes the representations of both users and items at the same time to minimize the average prediction error \cite{koren_2009}.
The predicted scores per exercise can be computed as $\hat{S} = U E$, where $\hat{S} \in \mathbb{R}^{n\times m}$ is the user-exercise score matrix, $U \in \mathbb{R}^{n\times l}$ contains the user's latent factors and $E \in \mathbb{R}^{l \times m}$ the exercise's latent factors, and $n=\text{number of users}$, $m=\text{number of exercises}$, $l=\text{number of latent factors}$.
The predicted exercise score for user $u$ and  exercise $e$ can then be computed as:
\begin{equation}
 \hat{s}_{ue} = \sum_{f \in [1:l]} U_{uf} E_{fe}
\end{equation}
This ensures that our representations for users and exercises are consistent with each other.
The problem then consists in optimizing the objective function
\begin{equation}
    \argmaxB_{U, E} ||S - \hat{S}|| + \lambda_1 ||U|| + \lambda_2 ||E||
\end{equation}
where the last two terms provide regularization, and $\lambda_1$, and $\lambda_2$ are hyperparameters.
To perform the optimization, we use the gradient descent algorithm in its convenient implementation in the deep learning library PyTorch\cite{pytorch}. 

In practice, however, where most users only complete a single course with a relatively small number of exercises contained in a limited time period, 
the standard MF approach described above faces the problem that for any two separate time periods the overlap of users and exercises is small. Which amounts to a kind of continuous cold start problem\cite{lika_2014}. 
To tackle it we propose \textit{dynamic matrix factorization}, which 
dynamically extends the user-exercise matrix
each time a new user or exercise appears in the interaction stream (adding a new row or column with values corresponding to an average exercise or user, respectively). We perform a small number of gradient descent training steps after each new observed interaction.  
This has three important consequences:
\begin{itemize}
    \item It allows us to derive representations for new users and new exercises already after a single recorded interaction, effectively solving the continuous cold start problem;
    \item Additional observations will help to gradually improve the representations over time; 
    \item Information learned from past users-exercise interactions can continuously flow into the representations of new users and exercises.
\end{itemize}
Our experiments confirm that representations indeed improve over time by comparing the performance of downstream models (see Sect. \ref{sec:successdropout}) at different numbers of user interactions.

\subsection{Success and dropout prediction} 
\label{sec:successdropout}
To model a user's interaction when presented with a new exercise, we need to predict her/his performance on the  exercise as well as her/his decision to drop out, i.e. not to finish the exercise, and quite the course. 

We use machine learning to predict these outcomes from the historical data.
As the input for the score prediction model, we concatenate the following into a single vector:
\begin{equation}
    {u_t, e_{t-n}, s_{t-n}, ..., e_{t-1}, s_{t-1}, e_t}
\end{equation}
where $u_t$ and $e_t$ are the user and exercise representations learned in Sect. \ref{sec:MF}, and $s_{t}$ is the performance score.
The performance of different machine learning algorithms is shown in Tab. \ref{sample-table}.
\begin{table}
  \caption{Performance comparison for different success  prediction model variants}
  \label{sample-table}
  \centering
  \begin{tabular}{llll}
    \toprule
    Model type    & Number of last exercises used  & RMSE \\
    \midrule
    Random Forest & 3   &  0.238  \\
    SVM    & 3  & 0.246   \\
    XGBoost  & 3  & 0.243 \\
    Random Forest & 10   &  \textbf{0.227}   \\
    SVM    & 10  & 0.237   \\
    XGBoost  & 10  & 0.239 \\
    \bottomrule
  \end{tabular}
\end{table}
The random forest model using the representations of the last $n=10$ exercises as input was selected for down stream tasks. 

\begin{figure}
  \begin{center}
    \includegraphics[width=0.55\textwidth]{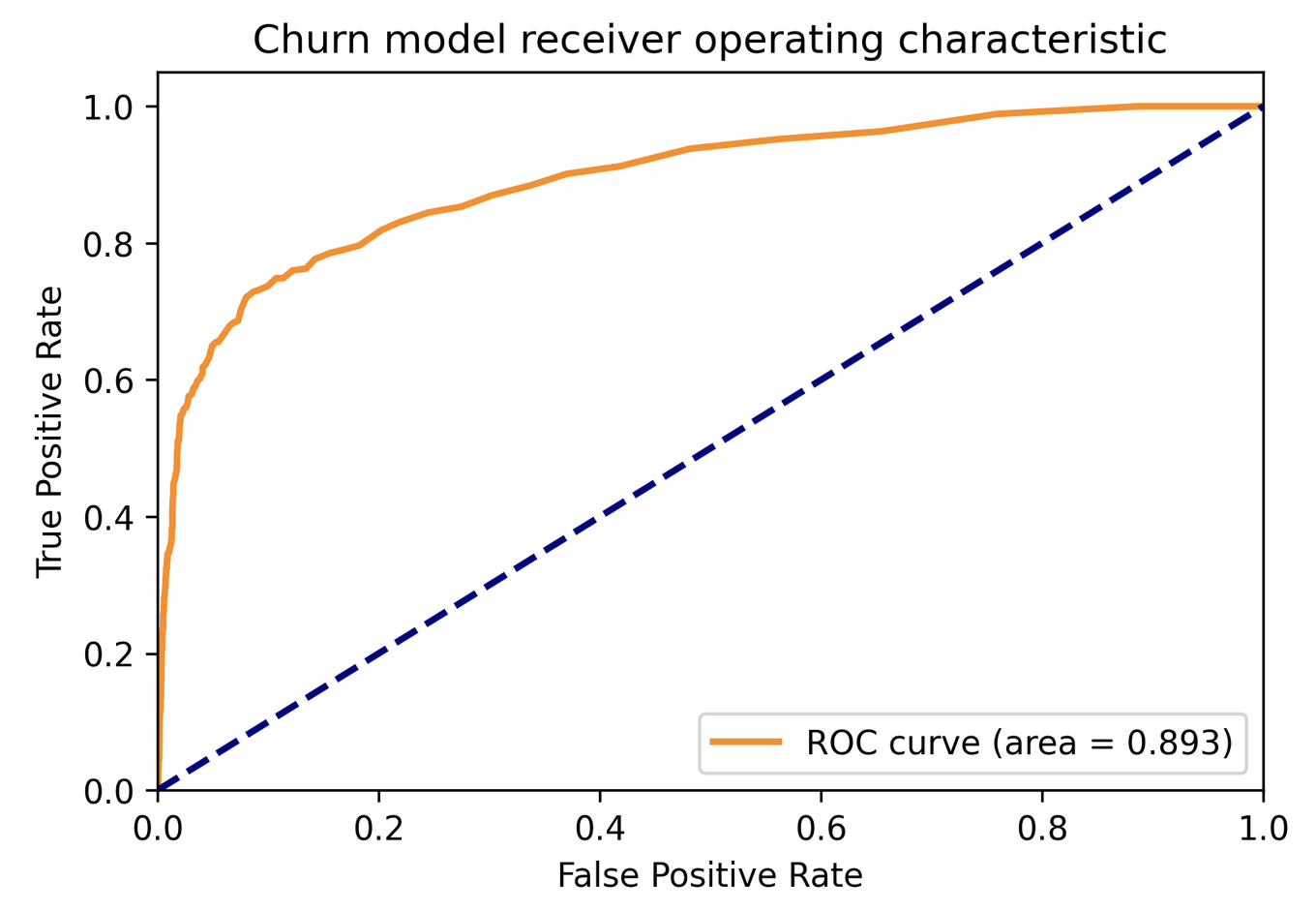}
  \end{center}
  \caption{Performance of dropout prediction model, measured by the ROC curve\cite{fawcett_2006_introduction}.}
  \label{fig:ROCcurve}
\end{figure}

As the input for the dropout prediction model, we concatenate the following into a single vector:
\begin{equation}
    {u_t, e_{t-n}, s_{t-n}, ..., e_{t-1}, s_{t-1}, e_t, s_t}
\end{equation}
Up-sampling of the minority class was used to deal with class imbalance. 
The performance of the dropout prediction model is shown by the ROC curve in Fig. \ref{fig:ROCcurve}.

\subsection{Student interaction simulator} 
\begin{figure}
  \begin{center}
    \includegraphics[width=0.65\textwidth]{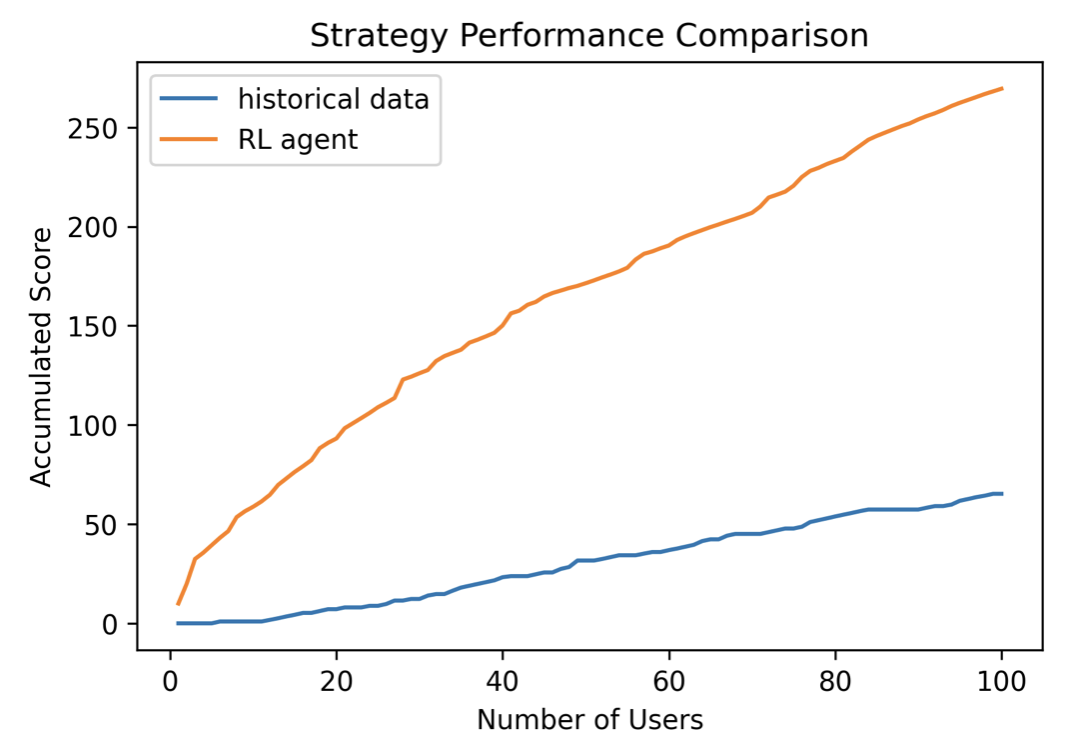}
  \end{center}
  \caption{Comparison of accumulated scores of episode rewards over time achieved if exercises are chosen like given in the historical data (blue) and by the RL agent (orange). In each episode, a new user is randomly selected.}
  \label{fig:performance}
\end{figure}

Given the models learned in Sect. \ref{sec:successdropout} it is now possible to construct a student interaction simulator. 
It receives the information about the next exercise the student should solve as input 
and returns a dropout decision as well as a performance score for the given exercise. 
It also returns a reward which can be used to train a potential learning agent tasked with selecting the next exercise to present to the student. 
In order to guide the student towards a desired target performance score $s_{\mathrm{target}}$ and minimize the dropout probability, $p_{\mathrm{dropout}}$, we define the reward function as: 
\begin{equation}
 R = \sum_{\mathrm{e}} \left[ (s_e - s_{\mathrm{target}})^2 + \alpha (1 - p_{\mathrm{dropout}}) \right]
\label{eq:reward}
\end{equation} 
where $s_e$ is the student's performance score for exercise $e$.
We compared the interaction with the simulation environment of two different strategies and observed the rewards they achieve
\begin{itemize}
    \item Applying the same exercise order as in the historical data. 
    \item Applying an AI-agent trained with PPO \cite{schulman_proximal_2017}.
\end{itemize}
The results are shown in Fig. \ref{fig:performance}. 
We observe that the AI agent is able to choose 
orders of exercises with consistently higher reward per user. This shows a clear potential for AI agents trained with our approach to improve learning outcomes in automated online courses.  

\section{Related work} 
Education offers a large number of opportunities for applications of artificial intelligence \cite{a7bfb7bf32544ebd8114cf324199317f}.
The advancement of digitization and 
learning management systems produces an increasing amount of data \cite{sciarrone_machine_2018}.
Analyzing it is the subject of the field of educational data mining \cite{romero_educational_2010, https://doi.org/10.1002/widm.1075}. 
E.g. it was found that dedicated feedback can enhance students' motivation \cite{ayouni_new_2021}. 
Machine learning (ML) has been applied in digital education in different ways, mostly using classical ML approaches \cite{munir_artificial_2022}. 
A big, mostly untapped potential for reinforcement learning exists in e-learning \cite{moubayed_e-learning_2018}.
Previous studies have mostly focused on student engagement \cite{ayouni_new_2021}, assuming that retention will follow. 
A number of studies focus on 
the prediction of student performance \cite{Anoopkumar2015ACS},
as well as the prediction and reduction of student dropout \cite{Tamada2019PredictingAR}.
The usefulness of recommendation systems for designing smart learning management system for digital learning has been shown in several studies, 
where the most popular methods are collaborative filtering and content-based filtering \cite{murad_recommendation_2018}. 
Recently, student profiling models have found increased attention, mostly using deep learning \cite{guan_artificial_2020}.
However, to date, no ML model exists which is able to actively steer the complete learning path of individual students in e-learning courses.

\section{Conclusions}
We developed a simulator for the automatic selection of exercises for individual students in automated online courses. We first use our dynamic matrix factorization method to derive neural representations for users and exercises, which provide the input features for success and dropout prediction models, trained via supervised machine learning. The resulting models could successfully be used in a simulator which enables the training of reinforcement learning agents. Initial tests show potential for such agents to optimize the learning path for each individual student in automated online  courses.
Next, we plan to incorporate the optimization of feedback, coding hints and assessments into the automated learning path
and test our AI agent in a real-world setting.

\section*{Acknowledgements} 
We thank Innosuisse for the support within 58831.1 INNO-ICT.



\bibliographystyle{plain}    
\bibliography{library} 

\end{document}